\documentclass[conference]{IEEEtran}
\IEEEoverridecommandlockouts

\usepackage{cite}
\usepackage{amsmath,amsfonts}
\usepackage{algorithmic}
\usepackage{graphicx}
\usepackage{textcomp}
\usepackage{xcolor}
\usepackage{multicol, multirow}
\usepackage{booktabs}
\usepackage{amssymb}
\usepackage{pifont}
\usepackage{caption}
\usepackage{subcaption}
\usepackage{enumitem}
\usepackage{url}

\def\BibTeX{{\rm B\kern-.05em{\sc i\kern-.025em b}\kern-.08em
    T\kern-.1667em\lower.7ex\hbox{E}\kern-.125emX}}

\newcommand{\xmark}{\text{\ding{55}}}

\begin{document}

\title{Enhancing In-the-Wild Speech Emotion Conversion with Resynthesis-based Duration Modeling
\thanks{This work was funded under the Excellence Strategy of the Federal Government and the Länder, and the project ``Mechanisms of Change in Dynamic Social Interaction'' (LFF-FV79, Landesforschungsförderung Hamburg).}
}

\author{\IEEEauthorblockN{Navin Raj Prabhu, Danilo de Oliveira}
\IEEEauthorblockA{\textit{Signal Processing} \\
\textit{University of Hamburg}\\
Hamburg, Germany \\
\{firstname.lastname\}@uni-hamburg.de}
\and
\IEEEauthorblockN{Nale Lehmann-Willenbrock}
\IEEEauthorblockA{\textit{Industrial and Organizational Psychology} \\
\textit{University of Hamburg}\\
Hamburg, Germany \\
nale.lehmann-willenbrock@uni-hamburg.de}
\and
\IEEEauthorblockN{Timo Gerkmann}
\IEEEauthorblockA{\textit{Signal Processing} \\
\textit{University of Hamburg}\\
Hamburg, Germany \\
timo.gerkmann@uni-hamburg.de}
}


\maketitle

\begin{abstract}

Speech Emotion Conversion aims to modify the emotion expressed in input speech while preserving lexical content and speaker identity. Recently, generative modeling approaches have shown promising results in changing local acoustic properties such as fundamental frequency, spectral envelope and energy, but often lack the ability to control the duration of sounds. To address this, we propose a duration modeling framework using resynthesis-based discrete content representations, enabling modification of speech duration to reflect target emotions and achieve controllable speech rates without using parallel data. Experimental results reveal that the inclusion of the proposed duration modeling framework significantly enhances emotional expressiveness, in the in-the-wild MSP-Podcast dataset. Analyses show that low-arousal emotions correlate with longer durations and slower speech rates, while high-arousal emotions produce shorter, faster speech.
\end{abstract}


\section{Introduction}
Speech is a fundamental social signal that plays a key role in enabling interactions, whether between humans or between humans and machines. It conveys essential information for the interaction, including lexical content, speaker identity, and expressed emotions \cite{schuller2018speech}. The task of speech generation and synthesis thereby is a crucial research topic in the fields of signal processing and human-computer interaction.
With the advent of generative deep neural networks, substantial improvements have been made in speech generation and synthesis \cite{prenger2019waveglow,hifigan,popov2021grad,jeong2021diff}. However, emotion-conditioned speech synthesis remains a significant challenge \cite{triantafyllopoulos2023overview, VCdu22c_interspeech, zhou2022emotionalESD, DurFlex}. In the context of human-computer interaction, the need for emotion-conditioned speech synthesis is crucial, not only to improve the naturalness and expressiveness of machine communication but also to enhance user engagement, foster empathy, and enable more effective and context-aware responses \cite{DurFlex, hifigan_rajprabhu23}.

Speech Emotion Conversion (SEC) is a sub-field of emotion-conditioned speech synthesis that aims to modify the emotion expressed in input speech while preserving lexical content and speaker identity \cite{triantafyllopoulos2023overview, zhou2022emotionalESD, metakreuk2022textless}. This requires precise control over prosodic attributes that convey emotional content, such as intonation, stress, rhythm, and loudness, which are controlled by the acoustic features of speech sounds, such as fundamental frequency, duration, energy, and spectral envelope. While it is appealing to control these attributes based on a target emotion, changing the corresponding acoustic feature for each prosodic component presents its own unique set of challenges \cite{DurFlex}. 

Generative deep neural networks, such as variational autoencoders (VAEs) \cite{Kingma2014}, generative adversarial networks (GANs) \cite{goodfellow2020generative}, and diffusion models \cite{song20score}, have been employed to the task of SEC with success in emotion conversion capabilities and improved naturalness in generated speech \cite{hifigan_rajprabhu23, prabhu2024emoconv}. However, these methods often overlook duration modeling in emotion conversion, resulting in inadequate control over crucial prosodic features such as rhythm and stress. Instead, they typically enforce fixed durations, where the emotion-converted output speech sounds have exactly the same duration as in the input, regardless of the intended emotion change. Interestingly, this is in contrast to the task of text-to-speech (TTS) synthesis, where duration modeling with duration-flexible speech generation is a common module, with proven improvements in the naturalness of synthesised speech \cite{popov2021grad}. 

Durflex-EVC \cite{DurFlex} introduced duration modeling in SEC with parallel data, where for each source utterance with a corresponding source emotion also a corresponding target utterance with a target emotion is available. Durflex-EVC learns discrete speech units from parallel target emotion speech and their repetitions. However, a particular challenge in duration modeling for emotion conversion arises when working with in-the-wild emotion datasets, as these lack parallel samples. As a result, there is no ground-truth duration reference for the target emotion, making accurate duration control more challenging. While in-the-wild datasets offer a richer and more naturalistic collection of emotional speech, along with greater speaker diversity and varied acoustic conditions \cite{salman24_interspeech, busquet2023voice, kzhou23_thesis}, their non-parallel nature limits their applicability for supervised duration modeling in emotion conversion \cite{hifigan_rajprabhu23, prabhu2024emoconv}. In this work, we aim to achieve duration modeling in SEC, focusing specifically on in-the-wild datasets without relying on parallel data.

In this work, we propose a resynthesis-based duration modeling approach to enhance SEC performance, which operates on discrete speech units and does not require parallel data. To enable duration modeling in a non-parallel setting, the proposed method is trained using a resynthesis setup, inspired by \cite{hifigan_rajprabhu23} and \cite{prabhu2024emoconv}.
In this setup, during training, the model simultaneously reconstructs the original input speech while the duration model learns to predict the repetition counts of discrete speech units. This prediction is based solely on the input speech, without any reliance on target speech. During inference, the trained duration model can predict the appropriate unit repetitions based on a target emotion embedding, enabling emotion-aware duration control. Experiments in the in-the-wild MSP-Podcast dataset show that the inclusion of the proposed duration modeling framework is beneficial for emotional expressiveness.


\section{Related Literature}
\subsection{Speech Emotion Conversion techniques}
SEC techniques can be broadly categorized into \textit{sequential} speech generation and \textit{parallel} speech generation models. Sequential generation models (e.g., \cite{robinson2019sequence, kim2020emotional, zhou2023_emointensitycontrol}) perform emotion-conditioned speech synthesis by sequentially generating speech units or frames, thereby achieving implicit duration modeling. However, they often face challenges such as difficulty in capturing long-term dependencies and high time complexity \cite{DurFlex}. This has motivated the development of parallel generation models  (e.g., \cite{vawgan_base, VCdu22c_interspeech, DurFlex}), which address these limitations by enabling parallel generation of speech frames. However, a key requirement of these models is the explicit modeling of the intended duration \cite{DurFlex}. 

Recently, there has been a shift in voice and emotion conversion research away from traditional scripted or acted-out speech, which often lacks the natural spontaneity of real-life conversations, towards the use of in-the-wild recorded speech \cite{hifigan_rajprabhu23, prabhu2024emoconv, salman24_interspeech}. Unlike acted-out data, which is essentially read-out speech, in-the-wild recordings are more spontaneous and capture diverse speaking styles, emotional expressions, nonverbal cues like laughter and lip smacks, and disfluencies such as repetitions, hesitations, and interruptions \cite{lotfian2017building, salman24_interspeech}. Empirical analyses using the NaturalVoices dataset \cite{salman24_interspeech} show that models trained on in-the-wild samples generate more natural and intelligible speech. However, such training requires methods that do not depend on parallel speech samples.

Raj Prabhu et al. \cite{hifigan_rajprabhu23} proposed a SEC framework using \textit{resynthesis} to eliminate the need for parallel data. A HiFiGAN-based vocoder reconstructs input speech from disentangled self-supervised learning (SSL)-based representations: discrete HuBERT embeddings for lexical content, speech emotion recognition (SER)-derived emotion embeddings, and speaker verification-based speaker embeddings. At inference, modifying the emotion embedding enables synthesis with the target emotion. Building on this, EmoConv-Diff \cite{prabhu2024emoconv} uses a diffusion decoder conditioned on “average-phoneme” mel features. While effective for in-the-wild SEC, these approaches lack duration modeling and cannot control speech rate based on the target emotion.

In this work, we use the resynthesis technique to achieve duration modeling under in-the-wild conditions without relying on any information from target emotion speech, neither target emotion durations nor speech embeddings. To the best of our knowledge, this is the first study to propose duration-flexible SEC that does not rely on parallel emotion speech samples.



\subsection{Duration Modeling techniques}

Duration Modeling in speech synthesis has been approached as a task of predicting the temporal alignment between lexical tokens (e.g., phonemes) and their respective acoustic features, essentially determining how long each unit should be held in the synthesised speech \cite{ren2019fastspeech, popov2021grad}. Modern neural TTS systems incorporate duration modeling either implicitly, using the attention mechanism \cite{tacotron, tacotron2}, or explicitly, using a duration predictor that predicts phoneme repetitions \cite{zeng2020aligntts, ren2019fastspeech, popov2021grad}. The explicit modeling approach has been preferred in non-autoregressive models like Grad-TTS \cite{popov2021grad} and FastSpeech \cite{ren2019fastspeech}, allowing for greater flexibility in modifying speaking style, emphasis, or speech rate.

Despite its demonstrated effectiveness in TTS, duration modeling has received limited attention in tasks like emotion and voice conversion. A likely reason for this omission is the difficulty of jointly training a duration model and learning to modify the prosodic features of the input speech during conversion. As a result, many emotion conversion models adopt a fixed-duration strategy, where the converted speech maintains the same duration as the input, regardless of the target emotion \cite{hifigan_rajprabhu23, prabhu2024emoconv}. 
This constraint limits the expressiveness of SEC systems by restricting their ability to adjust the timing of lexical units, and consequently, the rhythm and speech rate aligned with the intended emotional state.

DurFlex-EVC \cite{DurFlex} addresses the gap of incorporating duration modeling in SEC by using a so-called \textit{Unit Aligner} module to extract discrete content tokens and a \textit{Duration Predictor} to estimate their repetitions. However, this approach is not directly applicable to in-the-wild datasets, as it relies on speech units extracted from parallel target speech, which are unavailable in non-parallel settings. Additionally, the use of look-up table-based speaker and emotion embeddings further limits its adaptability to in-the-wild scenarios, where speaker and emotion conditions are more variable and less structured. Similarly, \cite{huangspkconv_icassp21} and \cite{wangemoconv_icassp25} also address duration modeling. While \cite{huangspkconv_icassp21} target speaker conversion and \cite{wangemoconv_icassp25} focus on emotion conversion with acted data and categorical emotion labels.

In this work, inspired by \cite{hifigan_rajprabhu23} and \cite{prabhu2024emoconv}, we propose a resynthesis-based duration modeling approach that is better suited for in-the-wild datasets. Our method relies solely on the input speech during training and does not require any information from the target emotion or target speech, making it fully compatible with non-parallel SEC tasks.


\section{Methodology}
\begin{figure}[t!]
\centering
\includegraphics[width=0.95\columnwidth]{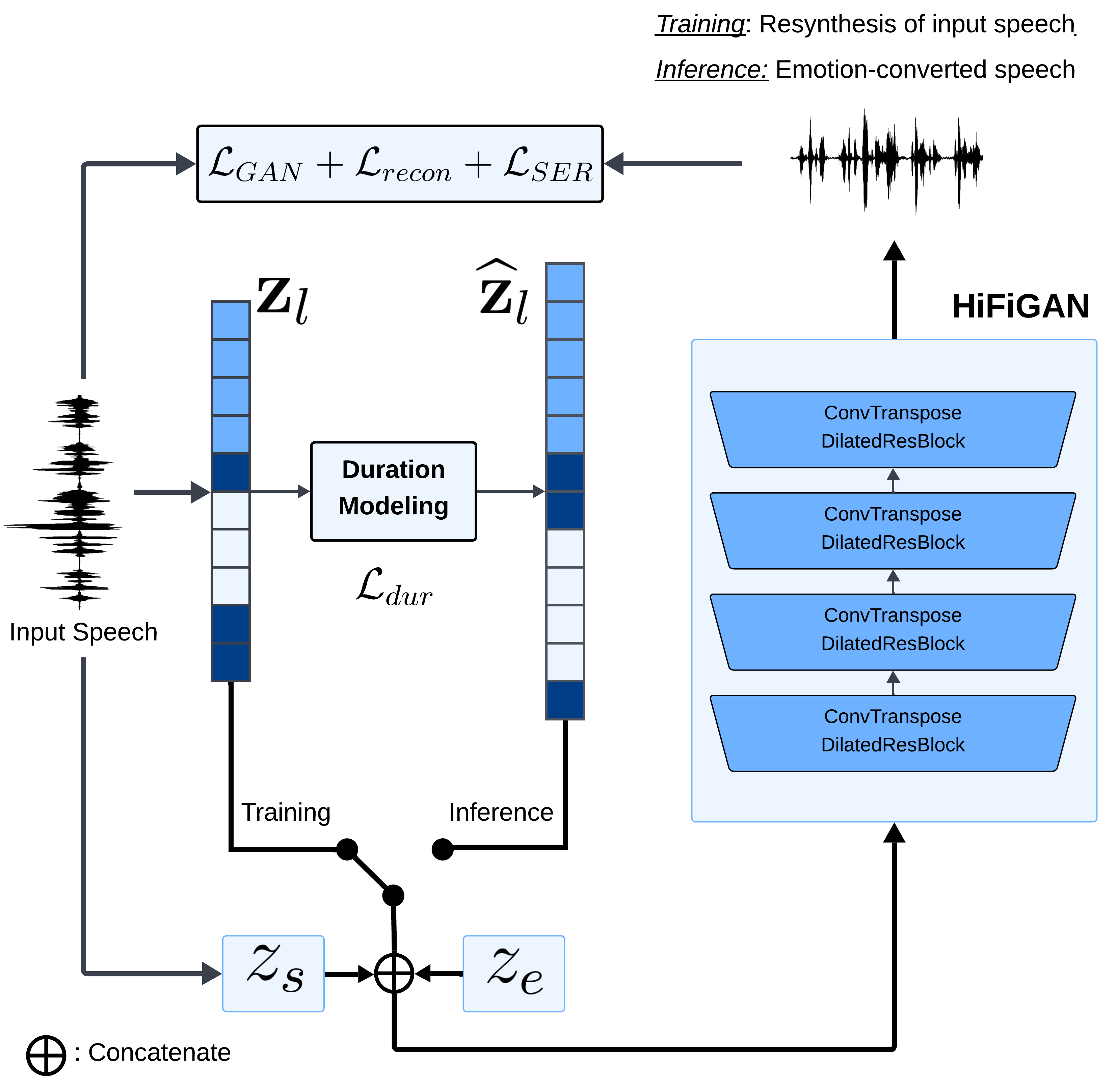}
	\caption{Overview of the speech emotion conversion framework.}
	\label{fig:sec_model}%
\end{figure}

The overall task of speech emotion conversion can be formulated as follows: given a single-channel audio input $\mathbf{x}_{l, s, e} \in \mathbb{R}^{1 \times T}$ representing a spoken utterance with lexical content $l$, speaker identity $s$, and annotated emotion level $e$, where the raw waveform is denoted as a sequence of samples $\mathbf{x} = [x_1, ..., x_T]$, the goal is to generate $\widehat{\mathbf{y}}_{l, s, \Bar{e}} \in \mathbb{R}^{1 \times T'}$, with $T'$ potentially different from $T$. This output should preserve the original lexical content $l$ and speaker identity $s$ from $\mathbf{x}_{l, s, e}$, while converting the expressed emotion to a desired target level $\Bar{e}$. With that intent, the length $T'$ is jointly modeled and the generated output is expected to be duration-flexible with respect to the desired target emotion $\Bar{e}$. 
We adopt the SSL-based HiFiGAN model from \cite{hifigan_rajprabhu23} as the SEC backbone for integrating our resynthesis-based duration modeling approach. Its original design, which is also trained using a resynthesis paradigm, makes it particularly well-suited for this purpose. The overall SEC methodology is depicted in Figure~\ref{fig:sec_model}.

\subsection{Disentangled Representations}

For the disentangled SSL-based representations input to the HiFiGAN decoder, we use the following encoded features:
\begin{enumerate}[label=(\roman*), align=left]
    \item \textit{Lexical representation} ($\mathbf{z}_l\in\mathbb{N}^{1\times N}$): Following \cite{metakreuk2022textless, hifigan_rajprabhu23, DurFlex}, we use discrete HuBERT units obtained via $k$-means clustering on continuous HuBERT features. Formally, $\mathbf{z}_l = [z_1, ..., z_N]$, where each $z_i$ is a positive integer and $N$ is the length of the input discrete unit sequence, corresponding to the number of frames in HuBERT's representations. Prior studies \cite{polyak21_interspeech_resynth, hsu2023revise, deoliveira2023leveraging} have shown that these units strongly correlate with the phonemic content of the utterance. The feature rate of these speech units is 49Hz.
    \item \textit{Speaker representation} ($z_s \in \mathbb{R}^{512}$): Adopted from \cite{hifigan}, we use a $d$-vector extracted from a pretrained WavLM-based speaker verification model \cite{Chen2021WavLM}.
    \item \textit{Emotion representation} ($z_e \in \mathbb{R}^{128}$): A continuous embedding obtained by applying a trainable linear transformation to the emotion label $e$ during training, and to the target emotion label $\bar{e}$ during inference.
\end{enumerate}

Unlike $\mathbf{z}_l$, both $z_s$ and $z_e$ are global utterance-level representations. To align them with the frame-level $\mathbf{z}_l$, we broadcast $z_s$ and $z_e$ across frames/discrete units, resulting in $\mathbf{z}_s$ and $\mathbf{z}_e$.

\begin{figure}[t!]
\centering
\includegraphics[width=\columnwidth]{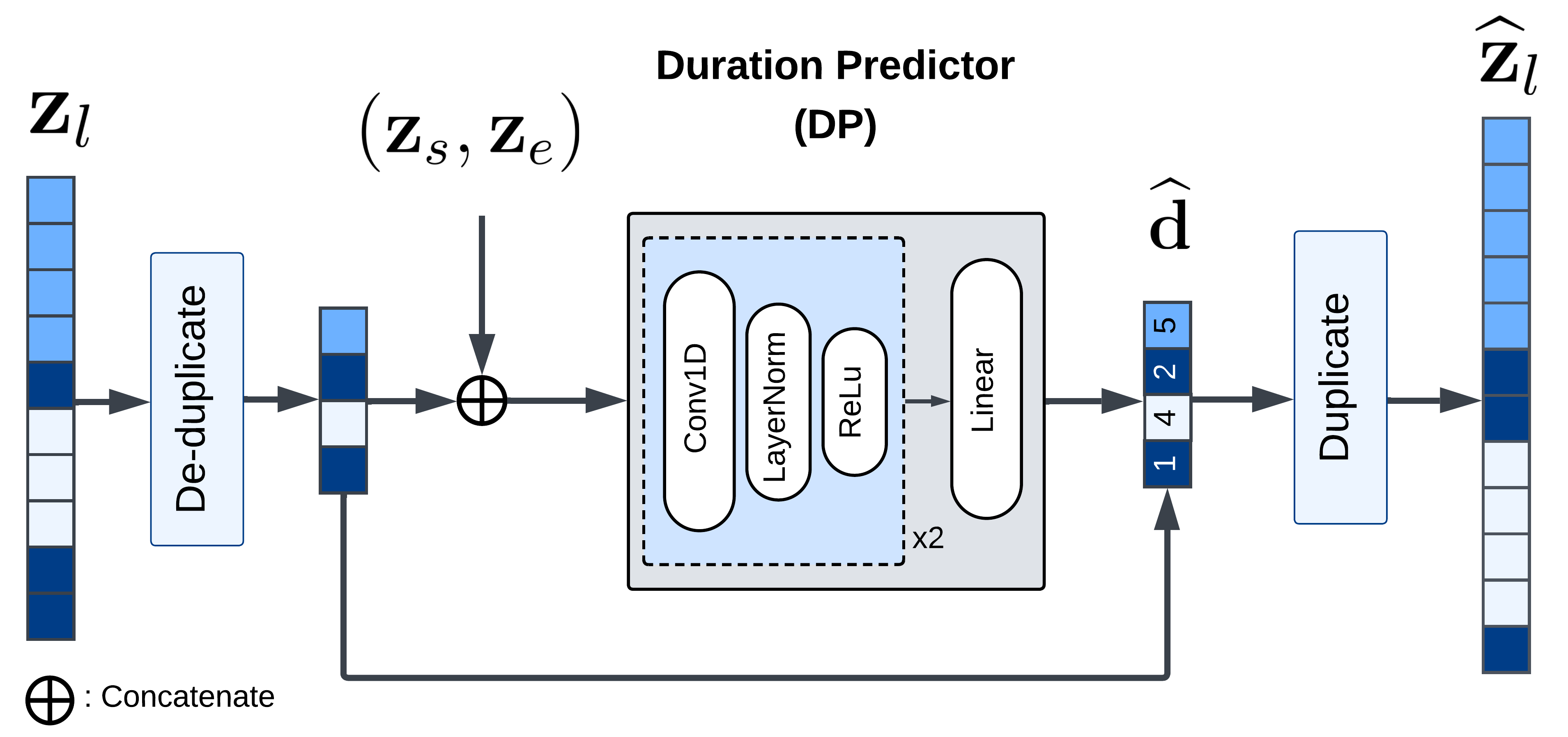}
	\caption{Overview of the duration modeling technique.}
	\label{fig:dur_model}%
\end{figure} 

\subsection{Duration Modeling}

An overview of the duration modeling technique can be seen Figure~\ref{fig:dur_model}. Based on $\mathbf{z}_s$ and $\mathbf{z}_e$, we perform duration modeling on the discrete HuBERT speech units $\mathbf{z}_l$, which represent the lexical content of input speech. Formally, we formulate the resynthesis-based duration modeling as follows: for $\mathbf{z}_l$ of input speech, we train a \textit{Duration Predictor} (DP) to predict the consecutive repetition of discrete speech units $\mathbf{d}$, conditioned on emotion and speaker representations. These repetitions represent the durations of each lexical unit.

Firstly, the frame-level $\mathbf{z}_l$ is de-duplicated to extract unit-level speech units, where the repetitions are ignored to obtain consecutive unique speech units. Secondly, this unit-level representation is fed as input to the predictor DP. To further achieve speaker and emotion conditioned duration modeling, we concatenate the speaker and emotion embeddings $(\mathbf{z}_s, \mathbf{z}_e)$ and pass them as an additional input to the predictor. 

The predictor is a simple deterministic neural network comprising two convolution layers and a linear layer to predict $\mathbf{d}_i$ for respective unit-level speech units, where $i$ is the index in the de-duplicated sequence of discrete speech units. As an example, if the speech units in $\mathbf{z}_l$ are $[1,1,2,2,2,1,3,3,3,3]$, the de-duplicated sequence would be $[1,2,1,3]$, and the target $\mathbf{d}$ would be $[2,3,1,4]$.
For stable training and to better account for outliers in durations, we predict durations in the $\log$-scale: $\log(\mathbf{d})$, as suggested in \cite{popov2021grad}. 
During training, the predicted $\log$-scale durations/repetitions are directly used in the loss function and the true frame-level $\mathbf{z}_l$ is used as the input to the HiFiGAN decoder. However, during inference, the predicted log durations $\widehat{\log (\mathbf{d}})$ are reversed back into duration units as follows:
\begin{equation}
    \widehat{\mathbf{d}} = \min\left(1, e^{\widehat{\log (\mathbf{d}}) + 1} \right).
\end{equation}
Finally, the reversed durations $\widehat{\mathbf{d}}$ are used to duplicate the unit-level speech units to obtain the duration modeled discrete lexical units $\widehat{\mathbf{z}}_l$. Note that, as per the resynthesis paradigm, we use the estimated $\widehat{\mathbf{z}}_l$ only during inference, and during training the true $\mathbf{z}_l$ is used. The final input to the HiFiGAN decoder is the combined concatenated representation: $(\mathbf{z}_l, \mathbf{z}_s, \mathbf{z}_e)$ during training and $(\widehat{\mathbf{z}}_l, \mathbf{z}_s, \mathbf{z}_e)$ at inference time.

\subsection{Loss Functions}
The overall training of the SEC architecture involves four different loss terms: (i) the adversarial based HiFiGAN loss $\mathcal{L}_{GAN}$, which is the same as used in \cite{metakreuk2022textless} and \cite{hifigan_rajprabhu23}, (ii) a reconstruction loss,
\begin{equation}
 {\mathcal{L}}_{recon}(G) = \sum_{\mathbf{x}} ||\phi(\mathbf{x}) - \phi(\hat{\mathbf{y}})||_1,   
\end{equation}
where $\phi$ is a function computing Mel-spectrogram, (iii) a speech emotion recognition loss which is used to better condition the SEC model on the emotion of input speech,
\begin{equation}
 \mathcal{L}_{SER} = \sum_{\mathbf{x}} [1 - {L}_{ccc}(e, E_{SER}(\hat{\mathbf{y}}))],
\end{equation}
where ${L}_{ccc}$ is the concordance correlation coefficient (CCC) \cite{lin_concordance_1989} computed between the ground-truth emotion $e$ of input speech, and the predicted emotion for resynthesised speech $E_{SER}(\hat{\mathbf{y}})$, and finally, (iv) the duration modeling loss $\mathcal{L}_{dur}$. We use the speech emotion recognition (SER) model introduced in \cite{wagner2023dawn} as the emotion predictor $E_{SER}(.)$. The emotion predictor is a wav2vec-based neural network trained on the MSP-Podcast dataset to predict the arousal of input speech.


As the duration modeling loss $\mathcal{L}_{\text{dur}}$, we experiment with three different loss functions, all computed on the logarithm of the ground-truth durations. Let the predicted log-durations be $\widehat{\mathbf{d}} = (\widehat{d}_1, \widehat{d}_2, \ldots, \widehat{d}_U)$, and the ground-truth durations be $\mathbf{d}~=~(d_1, d_2, \ldots, d_U)$, where $\tilde{d}_u = \log d_u$. Specifically, the four loss functions are: (i) \textit{mean squared error} ($\mathcal{L}_{mse}$), (ii) \textit{mean absolute error} ($\mathcal{L}_{abs}$), and (iii) \textit{uncertainty-based negative log-likelihood} (NLL) Loss, assuming a Gaussian distribution over log durations with predicted mean $\widehat{d}_u$ and predicted standard deviation $\sigma_u$ ($\mathcal{L}_{\text{NLL}}$).

Finally, the overall speech emotion conversion loss of the architecture is as follows:
\begin{equation}
    \mathcal{L}_{SEC} = \lambda_{1} \mathcal{L}_{GAN} + \lambda_{2} \mathcal{L}_{recon} + \lambda_{3} \mathcal{L}_{SER} + \lambda_{4} \mathcal{L}_{dur}, 
\end{equation}
where values of $\lambda_1$, $\lambda_2$ and $\lambda_3$ are adopted from \cite{hifigan_rajprabhu23}, and $\lambda_4$ is set to 2 after a grid-search based tuning.

\section{Experimental Setup}
\subsection{Dataset}
The dataset used in this study is the \emph{in-the-wild} MSP-Podcast dataset (v1.10) \cite{lotfian2017building}, which contains approximately $\approx$238hrs of audio sourced from podcasts, with utterance-level emotion annotations provided in terms of arousal, valence, and dominance. The dataset in contrast to predominant SEC datasets (e.g., ESD \cite{zhou2022emotionalESD}, IEMOCAP \cite{busso2008iemocap}) is larger, has utterances of variable duration, has over 1400 speakers, and contains naturalistic emotional expressions. For example, the ESD contains acted-out utterances from only 10 English speakers and only $\approx$29 hours of acted-out utterances. To the best of our knowledge, this is one of the few works to perform SEC on an in-the-wild dataset, along with \cite{hifigan_rajprabhu23} and \cite{prabhu2024emoconv}. 

For the purpose of this work, we focus exclusively on arousal annotations for SEC, following prior works on SEC under in-the-wild conditions \cite{hifigan, prabhu2024emoconv}. Performing SEC on the arousal dimension, instead of categorical representation has two advantages: (i) the circumplex-model based representation better captures the subtle difference between human emotion categories \cite{russell1980circumplex, rajprabhu_BBB_LDL}, and (ii) achieve implicit intensity control \cite{hifigan_rajprabhu23}, as opposed to an additional effort in the categorical representation case. 
The arousal annotations are rated on a 1–7 scale and exhibit a distribution with a mean $\mu = 4$ and standard deviation $\sigma = 0.95$. This indicates that samples are more concentrated in the mid-range (scores 3 to 5), with fewer examples at the extremes (scores 1 and 7). This skewed distribution mirrors the nature of emotional expression in real-world, in-the-wild scenarios, such as podcasts, where extreme emotional states are relatively rare.


\subsection{Validation Measures}
We evaluate the proposed methodology based on two key aspects: its speech emotion conversion (SEC) capabilities and the naturalness of the synthesised speech. To assess SEC performance, we use mean-squared error ($\mathcal{L}_{mse}$) and mean-absolute error ($\mathcal{L}_{abs}$), both computed between the target arousal $\Bar{e}$ and the SER model’s prediction on the resynthesised output, $E_{SER}(\hat{y})$.
For measuring the naturalness of the synthesised speech $\hat{y}$, we employ the wav2vec mean-opinion score (WVMOS) \cite{wvmos}, an objective speech quality metric derived from wav2vec2.0 \cite{baevski2020wav2vec}. WVMOS is fine-tuned on mean-opinion scores (ranging from 1 to 5) collected through listening tests from the 2018 Voice Conversion Challenge \cite{vcChallenge2018}, which focused specifically on naturalness. This makes WVMOS a suitable non-intrusive metric for evaluating the naturalness of $\hat{y}$.
It’s important to note that, since we do not use parallel data, we rely solely on non-intrusive evaluation metrics that do not require access to the ground-truth audio $\mathbf{y}_{l, s, \Bar{e}}$ for emotion conversion.

\begin{figure*}[t!]
     \captionsetup[subfigure]{justification=centering}
     \centering
     \begin{subfigure}[b]{0.48\textwidth}
            \includegraphics[width=\textwidth]{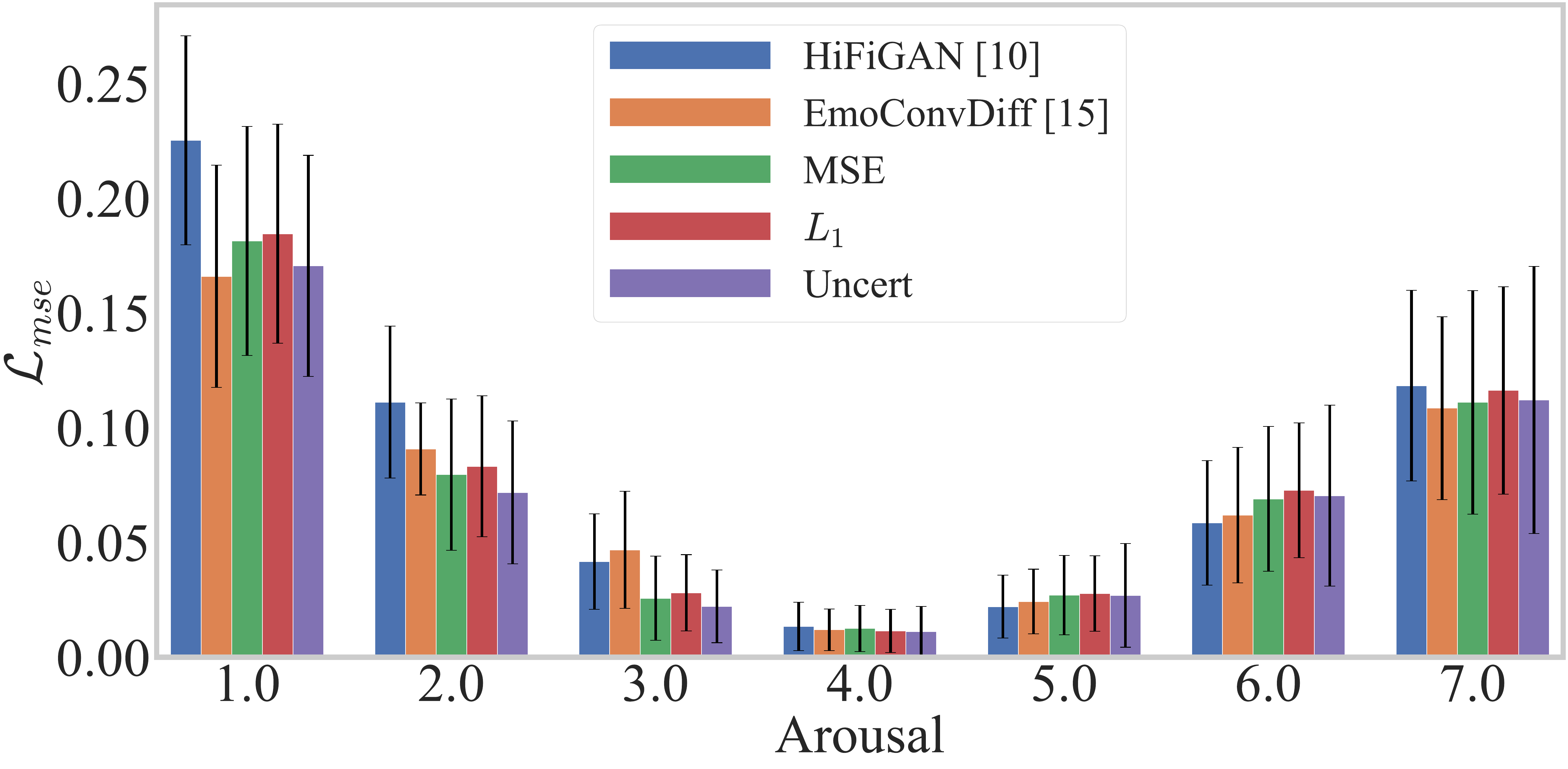}%
         \caption{SEC performance ($\mathcal{L}_{mse} \downarrow$).}
      \label{fig:sec-plot}%
     \end{subfigure}
     \hfill
     \begin{subfigure}[b]{0.48\textwidth}
            \includegraphics[width=\textwidth]{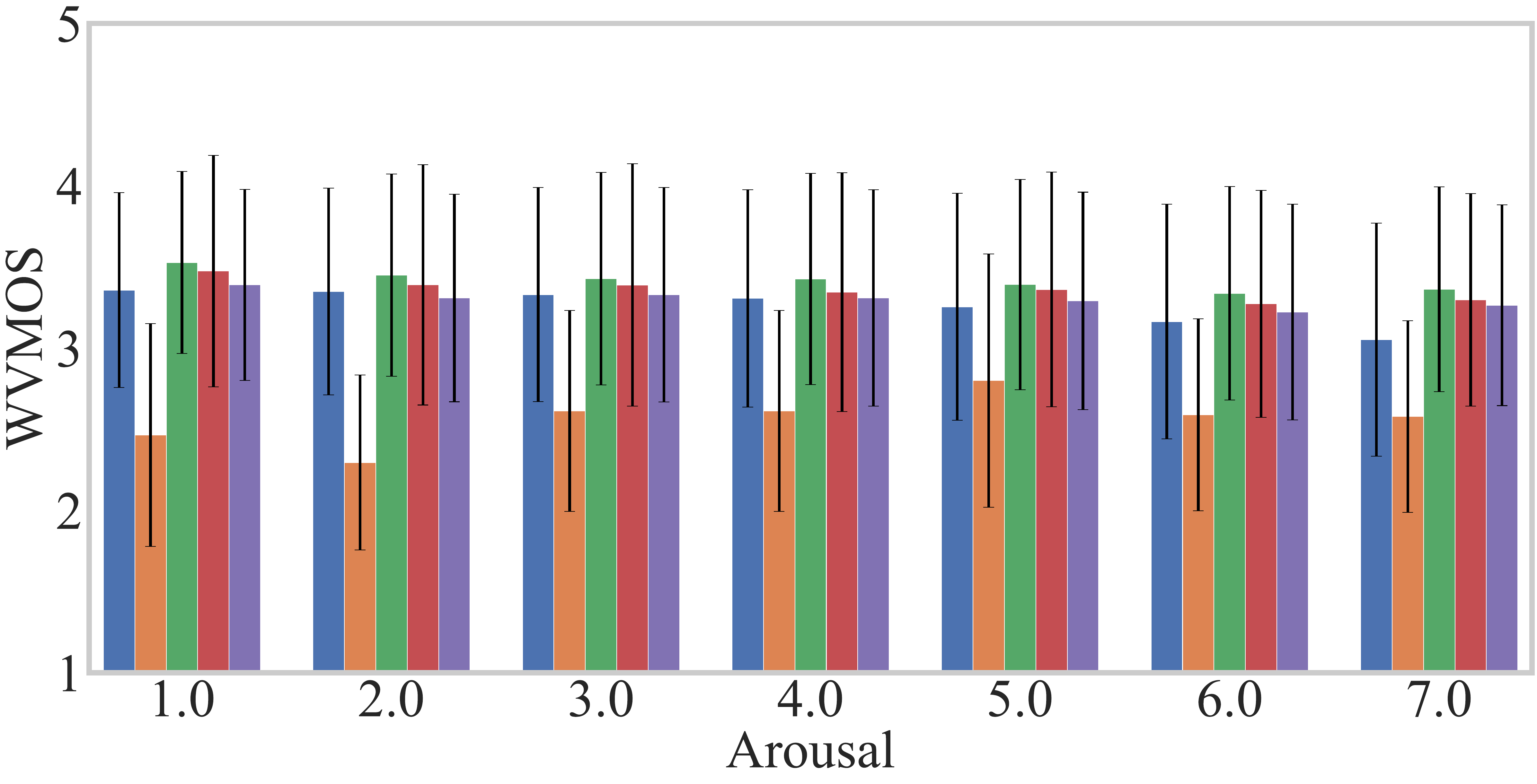}%
         \caption{WVMOS $\uparrow$ performance.}
      \label{fig:wvmos-plot}%
     \end{subfigure}
    
    \caption{Speech emotion conversion performance and naturalness of generated speech across arousal levels.}
    \label{fig:arous-level-wise-perf}
\end{figure*}
\begin{table}[]
     \centering
    \begin{tabular}{l|c|c|cc}
    \toprule
    \multirow{2}{*}{Model} & \multirow{2}{*}{\begin{tabular}[c]{@{}l@{}}DP\end{tabular}} & \multirow{2}{*}{WVMOS $\uparrow$} & \multicolumn{2}{c}{SER Error $\downarrow$} \\
     &  &  & $L_{mse}$ & $L_{abs}$ \\ \midrule
    HiFiGAN \cite{hifigan_rajprabhu23} & $\xmark$ & 3.26 & 0.084 & 24$\%$ \\
    EmoConv-Diff \cite{prabhu2024emoconv} & $\xmark$ & 2.56 & 0.072 & 21$\%$ \\ \midrule
    MSE & $\checkmark$  & \textbf{3.42} & 0.072 & 21$\%$ \\
    $L_1$ & $\checkmark$  & 3.36 & 0.075 & 22$\%$ \\
    $+$UnitAligner & $\checkmark$  & 3.16 & 0.086 & 27$\%$ \\
    Uncert & $\checkmark$ & 3.30 & \textbf{0.069} & \textbf{20$\%$} \\ \bottomrule
    \end{tabular}

    \caption{Overall performance of model versions. DP: Duration Predictor, $\checkmark$ indicates the respective model includes duration modeling, and $\xmark$ indicates it's absence.}
    \label{tab:performance}
\end{table}
\subsection{Model Versions}
As baselines for performance comparison, we use the HiFi-GAN \cite{hifigan_rajprabhu23} and EmoConv-Diff \cite{prabhu2024emoconv} architectures introduced earlier. Both are designed to handle in-the-wild data using the resynthesis paradigm that does not rely on parallel samples, similar to our proposed approach. Notably, neither model includes duration modeling, making them appropriate baselines for evaluating its impact. In fact, the HiFi-GAN architecture also serves as the backbone for our SEC model, into which we integrate duration modeling, further justifying its role as a baseline.
We evaluate four variants of the proposed model, each employing a different approach to duration modeling,
\begin{enumerate}[label=(\roman*), align=left]
    \item \textit{MSE}: trains the duration predictor using mean squared error (MSE) loss.
    \item \textit{$L_1$}: replaces MSE with mean absolute error.
    \item \textit{$+$UnitAligner}: integrates the Unit Aligner module from \cite{DurFlex}, which learns discrete speech units directly from data instead of relying on pretrained HuBERT units. These learned units are then utilized by the duration predictor, improving alignment between units and acoustic frames. With the inclusion of the Unit Aligner, this baseline corresponds to a reimplementation of DurFlex \cite{DurFlex} in our non-parallel, in-the-wild setting.
    \item \textit{Uncert}: introduces an uncertainty-aware duration predictor that estimates both the mean and variance of durations and is trained using Negative Log-Likelihood (NLL) loss for a probabilistic formulation.
\end{enumerate}

\section{Results}
\subsection{Influence of Duration Modeling}
The overall performance of the different versions of the proposed model, as compared to the baselines, is shown in Table~\ref{tab:performance}. From the results, we observe the following: Firstly, incorporating duration modeling into the HiFiGAN baseline leads to both increased naturalness in generated speech and enhanced speech emotion conversion capabilities. The MSE variant of the duration modeling attains a WVMOS of 3.42 and a $\mathcal{L}_{abs}$ of 21$\%$, representing an improvement over the HiFiGAN baseline, which achieves a WVMOS of 3.26 and a $\mathcal{L}_{abs}$ of 24$\%$. Secondly, it is evident that, except for the $+$UnitAligner version, all other duration modeling approaches consistently outperform the HiFiGAN baseline, highlighting the significance of duration modeling for SEC. A probable reason why the UnitAligner does not contribute to improved duration modeling is that it is better suited for training scenarios where parallel data samples are available, as was the case in the work that originally introduced it \cite{DurFlex}, and it does not provide additional benefit in a resynthesis-based training paradigm, where direct usage of HuBERT speech units $\mathbf{z}_l$ without alignment is more appropriate. Thirdly, we note that while duration modeling yields a considerable improvement over the HiFiGAN baseline, the gains over EmoConv-Diff are relatively small. This could potentially be attributed to differences in the decoder itself, as the diffusion-based decoder used by EmoConv-Diff is more complex and has already demonstrated improvements over HiFiGAN decoders in TTS tasks \cite{popov2021grad}.
Finally, among the duration modeling variants, the \textit{MSE} and \textit{Uncert} approaches emerge as the most effective. The \textit{MSE} variant yields slightly better naturalness, while the \textit{Uncert} variant achieves marginally better SEC performance. Overall, based on the empirical results, we recommend the \textit{Uncert} variant for duration modeling due to its strong SEC performance and improved naturalness over the baseline. The SEC capabilities can be further noted in the audio examples presented online\footnote{\url{https://sp-uhh.github.io/emoconv-gen}}.


\subsection{Performances across arousal levels}
In Figure~\ref{fig:arous-level-wise-perf}, the performance results are illustrated according to the target arousal level of the emotion-converted speech, considering both SEC capabilities (Fig.~\ref{fig:sec-plot}) and the naturalness of the generated speech (Fig.~\ref{fig:wvmos-plot}). Regarding SEC performance, the results in Fig.~\ref{fig:sec-plot} indicate that incorporating duration modeling proves particularly advantageous for generating low-arousal speech, with the duration modeling variants showing a noticeably larger improvement over the baseline at low arousal levels, and only a slight improvement at high arousal. Additionally, we observe that EmoConv-Diff achieves the best SEC performance for the extreme target arousal levels of 1 and 7, with the Uncert variant of duration modeling coming closest in performance.

From Fig.~\ref{fig:wvmos-plot}, we observe that duration modeling approaches consistently yield more natural-sounding speech compared to both the HiFi-GAN and EmoConv-Diff baselines. This underscores the importance of explicit duration modeling for enhancing speech naturalness. Although EmoConv-Diff demonstrates competitive SEC performance, it notably lacks in naturalness. This suggests a promising direction for future work: incorporating duration modeling into diffusion-based SEC methods such as EmoConv-Diff.

\subsection{Duration-flexible speech emotion conversion}\label{sec:dur_model}

\begin{figure}[t!]
\centering
\includegraphics[width=0.75\columnwidth]{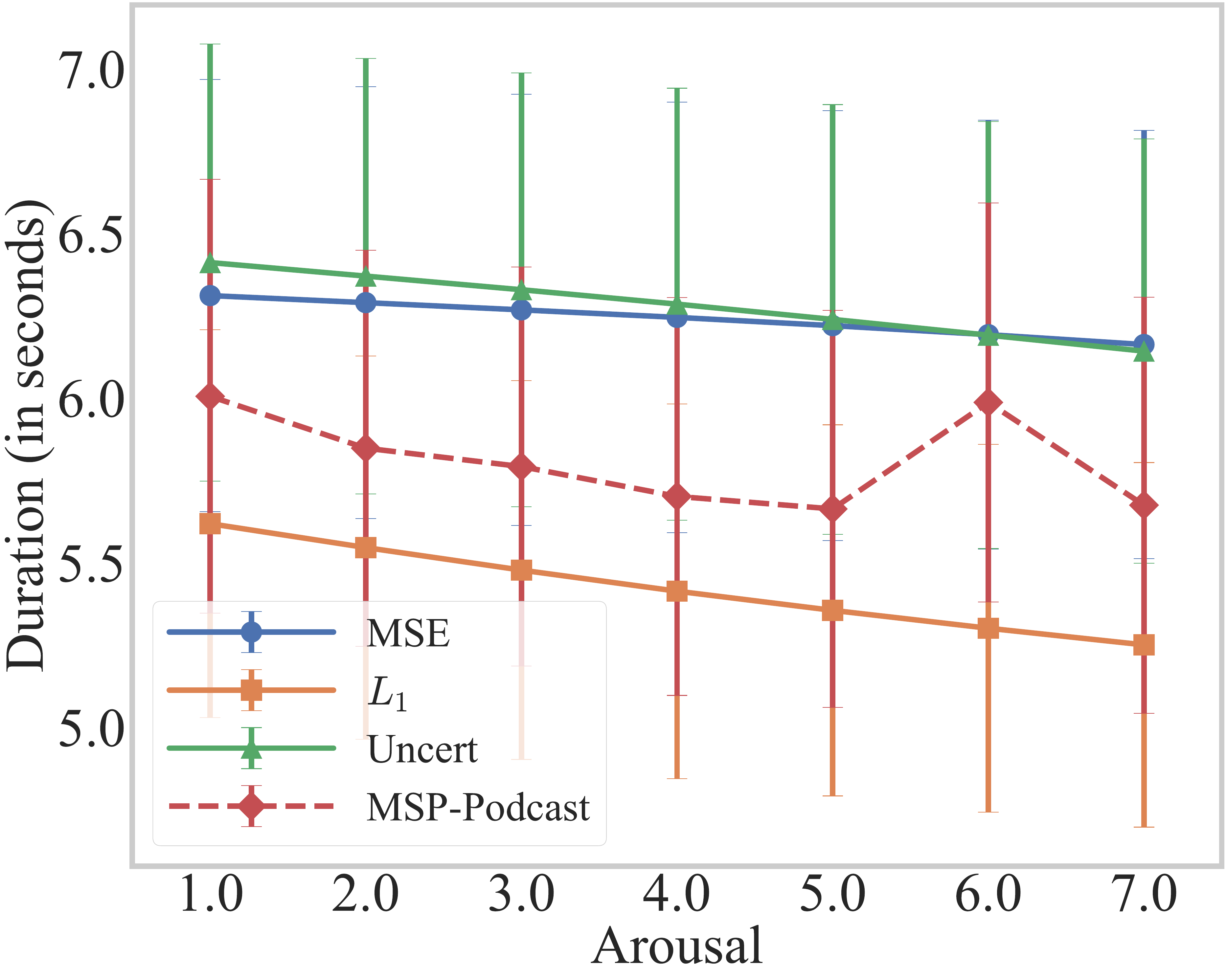}
	\caption{Mean and standard deviation of durations for emotion-converted speech across target arousal levels. The red dashed line represents the durations from the MSP-Podcast dataset.}
	\label{fig:dur_model_plot}%
\end{figure} 

To evaluate the effectiveness of duration modeling, Figure~\ref{fig:dur_model_plot} presents the mean durations (in seconds) and their standard deviations for emotion-converted speech generated by the various model variants. Additionally, we include the oracle mean durations from the MSP-Podcast dataset (denoted by the red dashed line) corresponding to the ground-truth arousal levels.
The figure clearly illustrates that all duration modeling variants successfully capture the inverse linear relationship between arousal level and speech duration: models tend to generate longer speech for low arousal levels and shorter speech for high arousal levels. This trend, also evident in the dataset reference line, is well reproduced by the SEC models incorporating duration modeling.

Among the different variants, the \textit{$L_1$} loss shows the most pronounced duration contrast, with the largest difference in mean duration between arousal level 1 and arousal level 7, measured as $\Delta_{1-7} = \text{0.37}$\,secs. Both the \textit{MSE} and \textit{Uncert} variants reflect similar patterns, with the \textit{Uncert} variant yielding a slightly higher $\Delta_{1-7}$ of $\text{0.21}$\,secs.
Overall, the \textit{$L_1$} variant tends to generate shorter duration speech compared to the other models. This behavior may stem from the nature of $L_1$ loss, being based on absolute error, is less sensitive to outliers (e.g., highly repetitive speech units). Consequently, it may underfit to high-repetition segments, treating them as noise and favoring shorter durations in general.

\subsection{Modification of prosody features}

\begin{figure}[t!]
\centering
\includegraphics[width=\columnwidth]{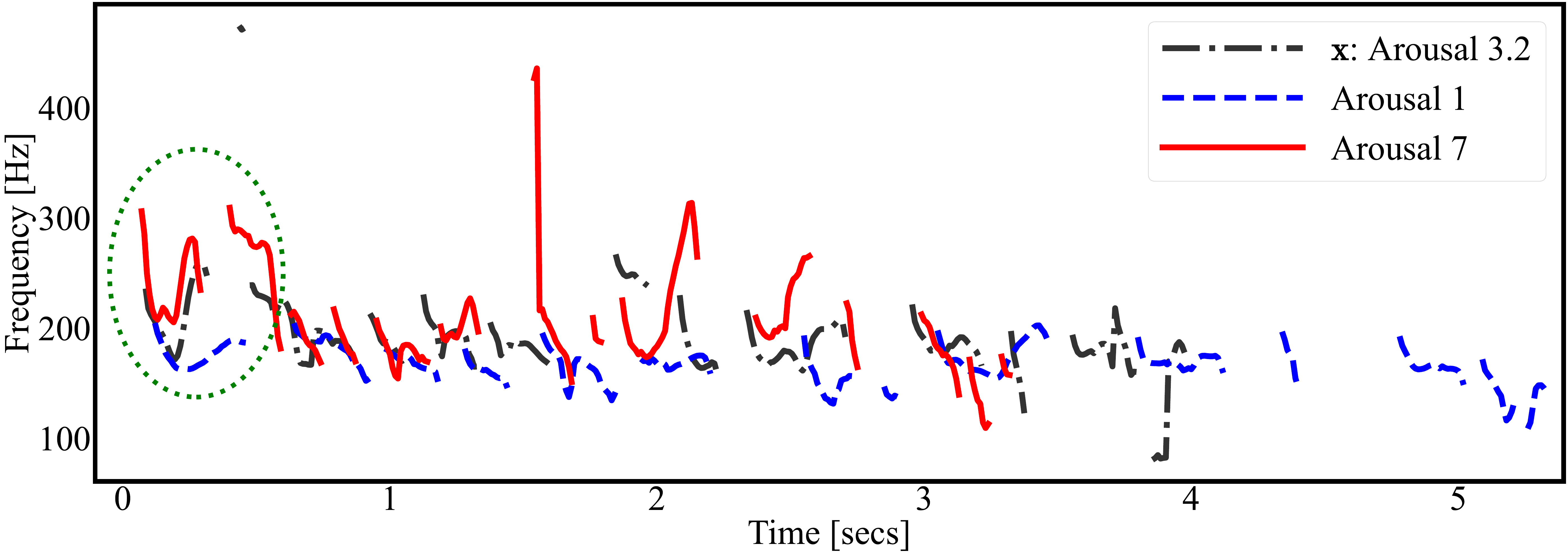}
	\caption{Pitch contours of the input speech $X$, along with emotion-converted speech for target arousal levels 1 and 7.}
	\label{fig:pitch_conts}%
\end{figure}

To examine the prosodic modifications achieved by the duration modeling-based SEC architecture, we present Figure~\ref{fig:pitch_conts}. It shows the pitch contours of the input speech (represented by black dashed lines), alongside the emotion-converted speech for target arousal level 1 (extreme low arousal, shown in blue) and arousal level 7 (extreme high arousal, shown in red). In the figure, a green dotted ellipse is used to highlight a region of interest. Due to the strong duration control demonstrated by the \textit{$L_1$} variant, as shown in Sec.~\ref{sec:dur_model}, the pitch contour analysis is conducted on speech generated by this variant.

The pitch contours in the figure reveal that the synthesised speech for high arousal ($\Bar{e} = 7$) exhibits a higher mean pitch and greater pitch variability compared to both the ground-truth speech ($e = 3.20$) and the synthesised speech for low arousal ($\Bar{e} = 1$). This observation is consistent with prior studies linking high emotional intensity to increased mean pitch \cite{zhou2023_emointensitycontrol}, and aligns with baseline research demonstrating effective pitch control. More importantly, we observe the impact of duration modeling, which yields a shorter duration for high arousal speech ($\approx$3.5\,secs), and a longer duration for low arousal speech ($\approx$5.3\,secs), compared to the ground-truth input speech of mid-level arousal ($\approx$ 4\,secs). Finally, within the highlighted region of interest (indicated by green dotted lines), it is evident that duration modeling enables effective control and modification of speech rate. Specifically, the high arousal speech, while exhibiting a higher mean and variability in pitch, also features a noticeably shorter voiced segment (red contour) than the corresponding voiced segment in the low arousal speech (blue contour).

\section{Conclusion}
In this work, we proposed a resynthesis-based duration modeling approach for speech emotion conversion that does not require parallel target speech samples—a key challenge due to the unavailability of ground-truth lexical durations during training. To overcome this, we employed a resynthesis training paradigm where the model learns to reconstruct input speech conditioned on lexical, emotion, speaker, and duration information. At inference time, emotion conversion is achieved by modifying the emotion embeddings.

We validated our approach on an in-the-wild dataset, evaluating both emotion conversion accuracy (using a pretrained SER model) and the naturalness of synthesised speech (via WVMOS). Pitch contour analysis confirms that our approach achieves not only pitch modulation but also speech rate control, producing shorter, faster speech for high arousal and longer, slower speech for low arousal. The results demonstrate the effectiveness of duration modeling, with consistent improvements in both SEC performance and naturalness over baseline methods.

\bibliographystyle{IEEEtran}
\bibliography{refs}

\def\ICASSP{IEEE Int. Conf. on Acoustics, Speech and Signal Proc. (ICASSP)}\def\INTERSPEECH{Interspeech}\def\ACII{IEEE Int. Conf. on Affective Comp. and Intelligent Interaction}\def\NIPS{Advances in Neural Inf. Proc. Systems (NeurIPS)}\def\ICML{Int. Conf. Machine Learning (ICML)}\def\ICLR{Int. Conf. on Learning Representations (ICLR)}\def\TSP{IEEE Trans. on Signal Proc.}\def\TAC{IEEE Tran. on Affective Computing}\def\TASL{IEEE Tran. on on Audio, Speech, and Language Processing}\def\WASPAA{IEEE Workshop on Applications of Signal Proc. to Audio and Acoustics (WASPAA)}\def\CVPR{IEEE/CVF Conf. on Computer Vision and Pattern Rec. (CVPR)}
\begin{thebibliography}{10}
\providecommand{\url}[1]{#1}
\csname url@samestyle\endcsname
\providecommand{\newblock}{\relax}
\providecommand{\bibinfo}[2]{#2}
\providecommand{\BIBentrySTDinterwordspacing}{\spaceskip=0pt\relax}
\providecommand{\BIBentryALTinterwordstretchfactor}{4}
\providecommand{\BIBentryALTinterwordspacing}{\spaceskip=\fontdimen2\font plus
\BIBentryALTinterwordstretchfactor\fontdimen3\font minus \fontdimen4\font\relax}
\providecommand{\BIBforeignlanguage}[2]{{%
\expandafter\ifx\csname l@#1\endcsname\relax
\typeout{** WARNING: IEEEtran.bst: No hyphenation pattern has been}%
\typeout{** loaded for the language `#1'. Using the pattern for}%
\typeout{** the default language instead.}%
\else
\language=\csname l@#1\endcsname
\fi
#2}}
\providecommand{\BIBdecl}{\relax}
\BIBdecl

\bibitem{schuller2018speech}
B.~W. Schuller, ``Speech emotion recognition: Two decades in a nutshell, benchmarks, and ongoing trends,'' vol.~61, no.~5.\hskip 1em plus 0.5em minus 0.4em\relax ACM New York, NY, USA, 2018, pp. 90--99.

\bibitem{prenger2019waveglow}
R.~Prenger, R.~Valle, and B.~Catanzaro, ``Waveglow: A flow-based generative network for speech synthesis,'' in \emph{ICASSP 2019-2019 IEEE International Conference on Acoustics, Speech and Signal Processing (ICASSP)}.\hskip 1em plus 0.5em minus 0.4em\relax IEEE, 2019, pp. 3617--3621.

\bibitem{hifigan}
J.~Kong, J.~Kim, and J.~Bae, ``Hifi-gan: Generative adversarial networks for efficient and high fidelity speech synthesis,'' \emph{\NIPS}, vol.~33, 2020.

\bibitem{popov2021grad}
V.~Popov, I.~Vovk, V.~Gogoryan, T.~Sadekova, and M.~Kudinov, ``Grad-tts: A diffusion probabilistic model for text-to-speech,'' in \emph{\ICML}.\hskip 1em plus 0.5em minus 0.4em\relax PMLR, 2021.

\bibitem{jeong2021diff}
M.~Jeong, H.~Kim, S.~J. Cheon, B.~J. Choi, and N.~S. Kim, ``Diff-tts: A denoising diffusion model for text-to-speech,'' \emph{arXiv preprint arXiv:2104.01409}, 2021.

\bibitem{triantafyllopoulos2023overview}
A.~Triantafyllopoulos, B.~W. Schuller, G.~{\.I}ymen, M.~Sezgin, X.~He, Z.~Yang, P.~Tzirakis, S.~Liu, S.~Mertes, E.~Andr{\'e} \emph{et~al.}, ``An overview of affective speech synthesis and conversion in the deep learning era,'' \emph{Proc. of the IEEE}, 2023.

\bibitem{VCdu22c_interspeech}
Z.~Du, B.~Sisman, K.~Zhou, and H.~Li, ``{Disentanglement of Emotional Style and Speaker Identity for Expressive Voice Conversion},'' in \emph{\INTERSPEECH}, Sep 2022.

\bibitem{zhou2022emotionalESD}
K.~Zhou, B.~Sisman, R.~Liu, and H.~Li, ``Emotional voice conversion: Theory, databases and esd,'' \emph{Speech Communication}, vol. 137, pp. 1--18, 2022.

\bibitem{DurFlex}
H.-S. Oh, S.-H. Lee, D.-H. Cho, and S.-W. Lee, ``Durflex-evc: Duration-flexible emotional voice conversion leveraging discrete representations without text alignment,'' \emph{\TAC}, 2025.

\bibitem{hifigan_rajprabhu23}
N.~Raj~Prabhu, N.~Lehmann-Willenbrock, and T.~Gerkmann, ``In-the-wild speech emotion conversion using disentangled self-supervised representations and neural vocoder-based resynthesis,'' in \emph{Proc. {ITG Conf. on Speech Comm.}}, Sep. 2023.

\bibitem{metakreuk2022textless}
F.~Kreuk, A.~Polyak, J.~Copet, E.~Kharitonov, T.~A. Nguyen, M.~Rivi{\`e}re, W.-N. Hsu, A.~Mohamed, E.~Dupoux, and Y.~Adi, ``Textless speech emotion conversion using discrete \& decomposed representations,'' in \emph{Proc. of the Conf. on Empirical Methods in Natural Language Processing}, 2022.

\bibitem{Kingma2014}
D.~P. Kingma and M.~Welling, ``{Auto-Encoding Variational Bayes},'' in \emph{\ICLR}, 2014.

\bibitem{goodfellow2020generative}
I.~Goodfellow, J.~Pouget-Abadie, M.~Mirza, B.~Xu, D.~Warde-Farley, S.~Ozair, A.~Courville, and Y.~Bengio, ``Generative adversarial networks,'' \emph{Communications of the ACM}, vol.~63, no.~11, pp. 139--144, 2020.

\bibitem{song20score}
Y.~Song and S.~Ermon, ``Improved techniques for training score-based generative models,'' in \emph{\NIPS}, 2020.

\bibitem{prabhu2024emoconv}
N.~Raj~Prabhu, B.~Lay, S.~Welker, N.~Lehmann-Willenbrock, and T.~Gerkmann, ``{EMOCONV-Diff}: Diffusion-based speech emotion conversion for non-parallel and in-the-wild data,'' in \emph{\ICASSP}, 2024, pp. 11\,651--11\,655.

\bibitem{salman24_interspeech}
A.~N. Salman, Z.~Du, S.~S. Chandra, İsmail Rasim~Ülgen, C.~Busso, and B.~Sisman, ``Towards naturalistic voice conversion: Naturalvoices dataset with an automatic processing pipeline,'' in \emph{Interspeech 2024}, 2024, pp. 4358--4362.

\bibitem{busquet2023voice}
F.~Busquet, F.~Efthymiou, and C.~Hildebrand, ``Voice analytics in the wild: Validity and predictive accuracy of common audio-recording devices,'' \emph{Behavior Research Methods}, 2023.

\bibitem{kzhou23_thesis}
K.~Zhou, ``Emotion modelling for speech generation,'' PhD thesis, National University of Singapore, 2022, available at \url{https://scholarbank.nus.edu.sg/handle/10635/243782}.

\bibitem{robinson2019sequence}
C.~Robinson, N.~Obin, and A.~Roebel, ``Sequence-to-sequence modelling of f0 for speech emotion conversion,'' in \emph{\ICASSP}, 2019, pp. 6830--6834.

\bibitem{kim2020emotional}
T.-H. Kim, S.~Cho, S.~Choi, S.~Park, and S.-Y. Lee, ``Emotional voice conversion using multitask learning with text-to-speech,'' in \emph{\ICASSP}, 2020, pp. 7774--7778.

\bibitem{zhou2023_emointensitycontrol}
K.~Zhou, B.~Sisman, R.~Rana, B.~W. Schuller, and H.~Li, ``Emotion intensity and its control for emotional voice conversion,'' \emph{\TAC}, 2023.

\bibitem{vawgan_base}
K.~Zhou, B.~Sisman, and H.~Li, ``Vaw-gan for disentanglement and recomposition of emotional elements in speech,'' in \emph{IEEE Spoken Language Tech. Workshop}, 2021.

\bibitem{lotfian2017building}
R.~Lotfian and C.~Busso, ``Building naturalistic emotionally balanced speech corpus by retrieving emotional speech from existing podcast recordings,'' \emph{\TAC}, vol.~10, no.~4, pp. 471--483, 2017.

\bibitem{ren2019fastspeech}
Y.~Ren, Y.~Ruan, X.~Tan, T.~Qin, S.~Zhao, Z.~Zhao, and T.-Y. Liu, ``Fastspeech: Fast, robust and controllable text to speech,'' \emph{\NIPS}, vol.~32, 2019.

\bibitem{tacotron}
Y.~Wang, R.~J. Skerry-Ryan, D.~Stanton, Y.~Wu, R.~J. Weiss, N.~Jaitly, Z.~Yang, Y.~Xiao, Z.~Chen, S.~Bengio, Q.~V. Le, Y.~Agiomyrgiannakis, R.~Clark, and R.~A. Saurous, ``Tacotron: Towards end-to-end speech synthesis.'' in \emph{\INTERSPEECH}, 2017, pp. 4006--4010.

\bibitem{tacotron2}
J.~Shen, R.~Pang, R.~J. Weiss, M.~Schuster, N.~Jaitly, Z.~Yang, Z.~Chen, Y.~Zhang, Y.~Wang, R.~Skerrv-Ryan, R.~A. Saurous, Y.~Agiomvrgiannakis, and Y.~Wu, ``Natural tts synthesis by conditioning wavenet on mel spectrogram predictions,'' in \emph{\ICASSP}, 2018, pp. 4779--4783.

\bibitem{zeng2020aligntts}
Z.~Zeng, J.~Wang, N.~Cheng, T.~Xia, and J.~Xiao, ``Aligntts: Efficient feed-forward text-to-speech system without explicit alignment,'' in \emph{\ICASSP}.\hskip 1em plus 0.5em minus 0.4em\relax IEEE, 2020, pp. 6714--6718.

\bibitem{huangspkconv_icassp21}
W.-C. Huang, Y.-C. Wu, and T.~Hayashi, ``Any-to-one sequence-to-sequence voice conversion using self-supervised discrete speech representations,'' in \emph{\ICASSP}, 2021, pp. 5944--5948.

\bibitem{wangemoconv_icassp25}
S.~Wang, T.~Qi, C.~Lu, Z.~Luo, and W.~Zheng, ``Enhancing zero-shot emotional voice conversion via speaker adaptation and duration prediction,'' in \emph{\ICASSP}, 2025.

\bibitem{polyak21_interspeech_resynth}
A.~Polyak, Y.~Adi, J.~Copet, E.~Kharitonov, K.~Lakhotia, W.-N. Hsu, A.~Mohamed, and E.~Dupoux, ``{Speech Resynthesis from Discrete Disentangled Self-Supervised Representations},'' in \emph{\INTERSPEECH}, 2021.

\bibitem{hsu2023revise}
W.-N. Hsu, T.~Remez, B.~Shi, J.~Donley, and Y.~Adi, ``Revise: Self-supervised speech resynthesis with visual input for universal and generalized speech regeneration,'' in \emph{Proc. of the IEEE/CVF Conf. on Computer Vision and Pattern Recognition}, 2023.

\bibitem{deoliveira2023leveraging}
D.~de~Oliveira, N.~Raj~Prabhu, and T.~Gerkmann, ``Leveraging semantic information for efficient self-supervised emotion recognition with audio-textual distilled models,'' in \emph{\INTERSPEECH}, 2023.

\bibitem{Chen2021WavLM}
S.~Chen, C.~Wang, Z.~Chen, Y.~Wu, S.~Liu, Z.~Chen, J.~Li, N.~Kanda, T.~Yoshioka, X.~Xiao, J.~Wu, L.~Zhou, S.~Ren, Y.~Qian, Y.~Qian, M.~Zeng, and F.~Wei, ``Wavlm: Large-scale self-supervised pre-training for full stack speech processing,'' \emph{IEEE Journal of Selected Topics in Signal Processing}, vol.~16, pp. 1505--1518, 2021.

\bibitem{lin_concordance_1989}
L.~I.-K. Lin, ``A {Concordance} {Correlation} {Coefficient} to {Evaluate} {Reproducibility},'' \emph{Biometrics}, vol.~45, no.~1, p. 255, Mar. 1989.

\bibitem{wagner2023dawn}
J.~Wagner, A.~Triantafyllopoulos, H.~Wierstorf, M.~Schmitt, F.~Burkhardt, F.~Eyben, and B.~W. Schuller, ``Dawn of the transformer era in speech emotion recognition: closing the valence gap,'' \emph{IEEE Tran. on Pattern Analysis and Machine Int.}, 2023.

\bibitem{busso2008iemocap}
C.~Busso, M.~Bulut, C.-C. Lee, A.~Kazemzadeh, E.~Mower, S.~Kim, J.~N. Chang, S.~Lee, and S.~S. Narayanan, ``Iemocap: Interactive emotional dyadic motion capture database,'' \emph{Language resources and evaluation}, vol.~42, no.~4, 2008.

\bibitem{russell1980circumplex}
J.~A. Russell, ``A circumplex model of affect.'' \emph{Journal of personality and social psychology}, vol.~39, no.~6, 1980.

\bibitem{rajprabhu_BBB_LDL}
N.~Raj~Prabhu, N.~Lehmann-Willenbrock, and T.~Gerkmann, ``End-to-end label uncertainty modeling in speech emotion recognition using bayesian neural networks and label distribution learning,'' \emph{\TAC}, vol.~15, no.~2, pp. 579--592, 2024.

\bibitem{wvmos}
P.~Andreev, A.~Alanov, O.~Ivanov, and D.~Vetrov, ``Hifi++: A unified framework for bandwidth extension and speech enhancement,'' in \emph{\ICASSP}, 2023, pp. 1--5.

\bibitem{baevski2020wav2vec}
A.~Baevski, Y.~Zhou, A.~Mohamed, and M.~Auli, ``wav2vec 2.0: A framework for self-supervised learning of speech representations,'' \emph{\NIPS}, vol.~33, pp. 12\,449--12\,460, 2020.

\bibitem{vcChallenge2018}
J.~Lorenzo-Trueba, J.~Yamagishi, T.~Toda, D.~Saito, F.~Villavicencio, T.~Kinnunen, and Z.-H. Ling, ``The voice conversion challenge 2018: Promoting development of parallel and nonparallel methods,'' in \emph{\INTERSPEECH}, Apr 2018.

\end{thebibliography}

\end{document}